\documentstyle[aps,prb,preprint]{revtex}
\begin{document}
\title{Theory of the Optical Conductivity in the
Cuprate Superconductors}
\author{Branko P.\ Stojkovi\'c}
\address{Department of Physics and Materials Research
Laboratory,\\
1110 West Green Street, University of Illinois, Urbana, IL 61801
}
\author{David Pines\cite{david-urbana}}
\address{Center for Nonlinear Studies\\
Los Alamos National Laboratory, Los Alamos, NM 87545}

\maketitle
\begin{abstract}
We present a study of the normal state optical conductivity in  
the cuprate superconductors using the nearly antiferromagnetic 
Fermi liquid (NAFL) description of
the magnetic interaction between their planar 
quasiparticles. We find that the  highly anisotropic 
scattering rate 
in different regions of the Brillouin zone, both as a function of 
frequency and temperature, a benchmark
of  NAFL theory, leads to an average relaxation rate of the
Marginal Fermi Liquid form for overdoped and optimally doped systems,
as well as for underdoped systems at high temperatures.
We carry out numerical calculations of 
the optical conductivity for several compounds
for which the input spin fluctuation parameters are known. 
Our results, which are in agreement with experiment on both overdoped
and optimally doped systems, show that NAFL theory explains
the anomalous optical behavior found in these cuprate
superconductors.
\end{abstract}

\vfill\eject
\section{Introduction}
\label{sec:intro}
\typeout{into on page \thepage}
Optical conductivity measurements show that  high
temperature superconductors
exhibit a number of anomalies when compared to 
the usual Drude-like
Fermi liquid (FL) behavior found in conventional 
metals.
Quite generally, experiment shows that 
the effective transport scattering rate, $1/\tau(\omega)$,
exhibits linear in $\omega$ behavior over a wide frequency range,
while the 
measurements of Puchkov et al\cite{timusk-review}  demonstrate that
significant deviations from this standard behavior occur,
especially  in underdoped materials at low temperatures,
in the so called pseudogap
regime, where $1/\tau(\omega)$ is strongly suppressed.
 These anomalies are 
often attributed to a possible non-Fermi liquid
ground state in these materials.\cite{anderson}

The linear frequency dependence of $1/\tau$ at relatively high
$\omega$ can  be regarded as the optical analogue of the  anomalous
resistivity, $\rho\propto T$, found in the superconducting cuprates,
and its presence has been considered as  major support for the 
 marginal-Fermi liquid (MFL) approach.\cite{varma} 
In the MFL approach, the self energy is given by:
\begin{equation}
\Sigma({\bf k},\omega) = \lambda \left(\omega \log
{x\over \omega_c} - i{\pi\over 2} x \,{\rm sign}\,
\omega\right) \label{eq:varma}
\end{equation}
where $x=\max(\vert \omega\vert,T)$, $\omega_c$ is a cutoff frequency and
$\lambda$  a
coupling constant, an expression which explained many of the early 
experiments.\cite{euro-varma}
However, Eq.\ (\ref{eq:varma}) is inconsistent with recent ARPES 
experiments\cite{zxshen} on the underdoped materials, which show that both 
the quasiparticle spectral weight and lifetime are highly anisotropic as 
one goes around the Fermi surface (FS), rather than being momentum 
independent as Eq.\ (\ref{eq:varma}) requires. 

Highly anisotropic 
quasiparticle behavior finds a natural explanation in the nearly 
antiferromagnetic Fermi liquid (NAFL) model of cuprates.\cite{pines-review}
in which the effective interaction between planar quasiparticles is 
assumed to be proportional to the dynamic spin susceptibility, 
$\chi(q,\omega)$, measured in NMR and INS experiments. Because 
$\chi(q,\omega)$ is strongly peaked in the vicinity of the 
antiferromagnetic wavevector 
${\bf Q}=(\pi,\pi)$,  quasiparticles on the FS 
located in the vicinity of the hot spots,\cite{tmrice} 
portions of the FS
which can be connected by wavevectors $\sim {\bf Q}$, interact strongly, 
while those located elsewhere (in {\em cold} regions)
interact comparatively weakly.

 NAFL theory, with spin-fluctuation parameters and a  quasiparticle 
spectrum taken from experiment, yields results consistent with low frequency  
NMR and INS experiments, as well as ARPES measurements of the Fermi 
surface properties.\cite{zxshen,chub-morr}
We have recently shown that it also provides a consistent qualitative, 
and in 
many cases quantitative, explanation of the measured changes with doping and 
temperature of both the longitudinal and Hall conductivities of the 
cuprates, and shown how one can extract the quite distinct 
lifetimes of ``hot'' and ``cold'' quasiparticles as a function of 
temperature directly from these measurements.\cite{sp-long}
In this paper we consider NAFL theory at 
infrared frequencies and show that it explains as well the measured 
anomalous optical behavior.

Our paper is organized as follows. Following a
review of the physical properties of NAFLs, we
discuss in Sec.\ \ref{sec:theory} both analytic and numerical calculations
of the highly anisotropic quasiparticle lifetime and the
quasiparticle relaxation rate at finite frequency $\omega$ and
temperature $T$. 
In Sec.\ \ref{sec:sigma_xx} 
we calculate the optical conductivity 
and discuss its limiting behavior. In Sec.\ \ref{sec:comparison}
we use  realistic band and spin-fluctuation
parameters to compare our results with experiment.
Our conclusions are presented in Sec.\
\ref{sec:conclusions}.

\section{The Physical Properties of NAFLs}
\label{sec:background}
\typeout{background on \thepage}

In the NAFL description of the normal state properties of the
superconducting cuprates, it is the magnetic interaction between
planar
quasiparticles which is responsible for the anomalous  spin and
charge
behavior. The magnetic properties of the system are specified by
the dynamical spin-spin response function of fermionic origin,
$\chi({\bf
q},\omega)$, which near a peak at a wavevector ${\bf Q}_i$
in the vicinity of 
${\bf Q}$, is assumed to take the mean-field form:\cite{mmp}
\begin{equation}
\chi(q, \omega) =   \frac{\chi_Q}{1 + ({\bf q} - {\bf
Q}_i)^2 \xi^2 - i
\omega/\omega_{sf} }
\label{eq:mmp}
\end{equation}
Hence 
$\chi_{Q}=\alpha \xi^2\gg \chi_0$ is the magnitude of the static spin
susceptibility at  ${\bf Q}_i$, $\xi$ is the antiferromagnetic correlation
length, $\omega_{sf}$ specifies the low frequency relaxational
mode,
brought about by the near approach to antiferromagnetism
and $\alpha$ is a temperature independent scale factor.
We use a system of units in which the lattice spacing $a=1$.
Although experiment suggests that the susceptibility
(\ref{eq:mmp})  quite generally possesses four peaks at the 
incommensurate wavevectors ${\bf Q}_i\approx {\bf Q}$,
since the introduction of the incommensuration does not introduce 
qualitatively new physics
 we shall assume for the most part that the spin
fluctuation spectrum possesses only a single peak at ${\bf Q}$.

The quasiparticle spectrum is assumed to take a tight-binding
form,
\begin{equation}
\epsilon_k=-2t( \cos k_x + \cos k_y) - 4t^\prime \cos k_x \cos
k_y-
2 t^{\prime \prime} (\cos (2k_x) + \cos (2k_y))
\label{eq:dispersion}
\end{equation}
where $t$, $t^\prime$ and $t^{\prime\prime}$ are the appropriate
hopping integrals, while the effective magnetic interaction
between the
planar quasiparticles is specified by
\begin{equation}
V_{eff}({\bf q},\omega)=g^2\chi({\bf
q},\omega).\label{eq:interaction}
\end{equation}
For a given system the parameters
$\chi_{{\bf Q}_i}$, $\xi$, $\omega_{sf}$ which determine
$\chi({\bf q},\omega)$ are
taken from fits to NMR and INS (inelastic neutron scattering)
experiments, while the effective coupling constant, $g$, is
assumed to
be momentum independent for  wavevectors near ${\bf Q}_i$. As
discussed by
Chubukov et al,\cite{CPS} the effective
interaction,
Eq.\ (\ref{eq:interaction}), can be, in principle, derived microscopically
starting
with, e.g., a one-band Hubbard model.

The effective interaction $V_{eff}$, Eq.\
(\ref{eq:interaction}),
has the obvious property that for
sufficiently large correlation lengths it is highly peaked for
momentum transfers in the vicinity of the antiferromagnetic
wavevector ${\bf Q}$. The consequences of this peaking are
hard to overestimate: if the FS of the system of fermions, defined by
the quasiparticle dispersion (\ref{eq:dispersion}), is such that it
intersects the magnetic Brillouin zone  (see Fig.\ \ref{fig:fs}), then
quasiparticles in the vicinity of these
intersection points on
the FS, the hot spots,\cite{tmrice}
 are much more strongly scattered by the spin-fluctuations than
those which are on other parts (cold regions)
of the FS. This leads to
strongly anisotropic quasiparticle behavior,
since the resulting temperature and frequency 
variation of the quasiparticle
scattering rates at and far away from hot spots is
in general very different.
We shall return to
this point in the following Section.

Barzykin and Pines (BP) have used Eq.\ (\ref{eq:mmp}) to
analyse NMR results in the cuprates.\cite{BP} 
They find that for underdoped systems the low frequency  magnetic
behavior possesses three distinct normal state phases,
and that the characteristic frequencies and lengths which enter Eq.\ 
(\ref{eq:mmp}) are connected by a dynamic scaling relationship,
$\omega_{sf}\propto \xi^{-z}$, in two of the phases. Thus, above 
the temperature, $T_{cr}$, at which the temperature dependent uniform
susceptibility, $\chi_0(T)$, possesses a maximum, the system exhibits 
non-universal mean field (MF)  behavior with
dynamical
exponent $z=2$; 
$\omega_{sf}\sim 1/\xi^2$, and the product,
$\chi_{\bf Q}\omega_{sf}\sim\omega_{sf}\xi^2$ is
independent of
temperature.
From a detailed analysis of the NMR experiments, Barzykin and Pines 
conclude that the crossover temperature, $T_{cr}$, is determined by the 
strength of the AF correlations, with $\xi(T_{cr})\approx 2$. Below 
$T_{cr}$, down to a second crossover temperature, $T_*$, underdoped systems 
exhibit non-universal, $z=1$, pseudoscaling (PS)
behavior. Thus for $T_*\leq T\leq T_{cr}$, it is
$\omega_{sf}\xi$ which is independent of temperature.
NMR experiments show that above $T_{cr}$, in the MF regime,
$\omega_{sf}$ and $1/\xi^2$
scale linearly with $T$, i.e., $\omega_{sf}=A+BT$, while in the PS regime,
between $T_*$ and $T_{cr}$,
it is $\omega_{sf}$ and $\xi^{-1}$ which scale linearly
with $T$, albeit with a possibly somewhat different slope and intercept
of $\omega_{sf}$ than
that found above $T_{cr}$. This behavior
has now been verified experimentally for the
YBa$_2$Cu$_3$O$_{7-x}$,  La$_{2-x}$Sr$_x$CuO$_4$ and 
YBa$_2$Cu$_4$O$_8$ systems,\cite{BP}
 as well as the 
Hg and 2212 BSCCO
compounds.\cite{joerg-nmr} Below $T_*$, in the
pseudogap (PG) regime, $\xi$ becomes independent of temperature
while $\omega_{sf}$, after exhibiting a minimum near $T_*$, rapidly
increases (roughly as $1/T$) as $T$ decreases toward $T_c$.  
These changes in the magnetic
fluctuation spectrum are accompanied by (indeed result from)
changes in the quasiparticle spectrum.\cite{CPS}
Thus, the PS regime is characterized by a
strong temperature variation of the quasiparticle spectrum,
resulting in a FS evolution\cite{CPS} 
which has non-trivial consequences for
the transport properties of underdoped systems.\cite{sp-long}
The crossovers seen in the low frequency
magnetic behavior thus possess, to a considerable extent, their charge
counterparts in transport experiments.

From a magnetic perspective, the so called optimally doped
systems (e.g., YBa$_2$Cu$_3$O$_{6.93}$ and
La$_{1.85}$Sr$_{0.15}$CuO$_4$) are a special case
of the underdoped systems, in which $T_*$ is comparatively close
to
$T_c$. Overdoped cuprates are
defined as
those for which $T_{cr}<T_c$. For these systems, then, 
the
antiferromagnetic correlations are comparatively weak,
with $\xi \leq 2$, 
$\chi_0(T)$ is at most weakly temperature dependent, while
$\omega_{sf}\propto \xi^{-2}$ follows the linear in $T$ behavior
found  in
the underdoped systems above $T_{cr}$. Examples of overdoped
systems
are the $T_c\sim 40$K Tl 2212 system and La$_{2-x}$Sr$_x$CuO$_4$
for
$x\geq 0.24$.

As Monthoux and Pines,\cite{MP-pseudo} and Chubukov et al\cite{CPS} have 
emphasized, the physical origin of the highly anomalous behavior displayed 
by the nearly antiferromagnetic Fermi 
liquids resides in the non-linear feedback of changes in the {\em hot} 
quasiparticle spectrum on the strong quasiparticle interaction, $\sim 
\chi({\bf q},\omega)$, which determines that {\em hot} quasiparticle 
behavior. For systems above $T_{cr}$, or in overdoped systems, that 
feedback is negative; strong coupling effects cause $\chi_0(T)$ to 
increase weakly as the temperature decreases, while an RPA or mean field 
description description suffices to determine the relationship between 
$\omega_{sf}$ and $\xi$. At a critical value of the strength of the AF 
correlations, that feedback becomes positive; a weak pseudogap develops 
in the hot 
quasiparticle spectrum, bringing about a linear decrease in $\chi_0(T)$ as 
$T$ decreases below $T_{cr}$, while the quasiparticle damping of 
spin excitations becomes $T$ dependent in such a way that $\omega_{sf}$ 
displays $z=1$ pseudoscaling behavior, with the strength of the AF 
correlations, $\xi$, growing as $(a+bT)^{-1}$. The second crossover 
temperature, $T_*$, marks the transition to ``strong'' pseudogap behavior: 
the AF correlations become frozen; $\omega_{sf}$ increases rapidly as $T$ 
decreases below $T_*$, while $\chi_0(T)$ falls off rapidly between $T_*$ 
and $T_c$. It is the interplay between these changes in the quasiparticle 
spectrum (seen directly in ARPES measurements) and the spin fluctuation 
spectrum, and effective quasiparticle interaction, seen directly in NMR and 
INS experiments, which is responsible for the anomalous quasiparticle 
lifetimes we now consider.

\section{Quasiparticle Lifetime at Finite Frequency}
\label{sec:theory}
\typeout{Section theory on Page \thepage}

We begin by estimating the effective scattering rate,
for quasiparticles near the FS,
as a function of frequency and temperature  
using  simple perturbation theory.
The quasiparticle lifetime is determined by the imaginary part of 
the single particle self-energy, which, 
for the effective interaction (\ref{eq:interaction}), in the 
second order Born approximation, reads:
\begin{equation}
Im \Sigma(k,\omega) = \sum_{k^\prime} 
g^2 \, {\rm Im}
\chi({\bf k-k^\prime},\omega-\epsilon^\prime) \,
[n_0(\epsilon^\prime-\omega) + f_0
(\epsilon^\prime)]\label{eq:scatkq}
\end{equation}
where 
$\epsilon^\prime\equiv
\epsilon_{k^\prime}$, and
$n_0(\epsilon)$,  and $f_0(\epsilon)$ are the Bose and Fermi
distribution functions respectively.
The transport relaxation rate is usually given by:
\begin{equation}
{1\over \tau_k(\omega)} = \int {d^2{k}'\over (2\pi)^2}
2g^2 \, {\rm Im}
\chi({\bf k-k^\prime},\epsilon^\prime-\omega) \,
[n_0(\epsilon^\prime-\omega) + f_0
(\epsilon^\prime)]
(1-\gamma_{k,k^\prime})\label{eq:tau_vertex}
\end{equation}
where $\gamma_{k,k^\prime}$ is a temperature independent 
vertex function which removes 
forward scattering processes
from the calculated transport relaxation rate. As we shall
see below, for overdoped,
optimally doped, and even underdoped systems above $T_*$, the dominant 
contribution to
$1/\tau_k$ comes from  large momentum transfer
scattering processes and hence the use of $\gamma_{k,k^\prime}$
for forward scattering corrections
plays  only a minor role. Therefore we assume
 $\gamma=0$, and  return later to the role of the vertex corrections 
at low temperatures in underdoped systems.

We make 
the usual change of variables in the integral in Eq.\
(\ref{eq:tau_vertex}):
\begin{equation}
\int d^2{k}'\rightarrow \int d\epsilon\prime \int {dk^\prime \over \vert
{\bf v}\vert}\label{eq:varchange}
\end{equation}
where $k^\prime$ is the component of momentum ${\bf k^\prime}$
parallel to the equipotential lines ($\epsilon^\prime=const$),
and carry out the integration over $\epsilon'$ for
${\bf k}$ near the FS and $\epsilon \ll t$, where $t$ is
the hopping matrix element.
On making the substitution $x=(\varepsilon^\prime-\omega)/T$ in
Eq.\ (\ref{eq:tau_vertex}), it becomes:
\begin{equation}
{1\over \tau_k(\omega,T)} = 2 \alpha g^2\omega_{sf}\xi^2
\int_{FS} {d{k}'\over 4\pi^2}
I({\bf k},{\bf k}',\varepsilon)\label{eq:tauI}
\end{equation}
where
\begin{equation}
I({\bf k,k^\prime},\varepsilon)=
\int dx\, \left[ {1\over \exp (x)-1} + 
{1\over \exp (x+\varepsilon)+1}\right]
\,{x\over \Omega^2 + x^2}
\label{eq:integral}
\end{equation}
and 
\begin{equation}
\varepsilon = \omega/T\qquad \Omega = \omega_{sf} [1 + \xi^2
({\bf k-k^\prime-Q})^2]/T\equiv
\omega_{k,k^\prime} /T \label{eq:def_eps}.
\end{equation}
Note that the definition (\ref{eq:def_eps}) 
 effectively removes
the temperature from the  calculation. 

Obviously, the integral $I$ is hard to solve  analytically. However
we can obtain its value in 
appropriate limits, and then map its
 behavior with a  universal function which possesses the
correct limiting values.

{\em Low Frequency Limit}, $\varepsilon\ll 1$:
In this limit  one can expand $1/[\exp(x+\varepsilon)+1]$
in Eq.\ (\ref{eq:integral}). The first non-vanishing correction to 
$I(\varepsilon=0)$
is  $I_2\varepsilon^2$, where 
\begin{equation}
I_2 = {1\over 2} \int_{-\infty}^\infty
dx {x\over x^2 + \Omega^2} {\partial^2 \over
\partial x^2}\left({1\over \exp(x)+1}\right).
\end{equation}
When $\Omega\ll 1$ one finds
$I_2 = 7\zeta(3)/2\pi^2 - \pi\Omega/8$ to lowest order in $\Omega$, while
in the limit $\Omega\gg 1$ one finds
$I_2\approx 1/\Omega^2$ to lowest order in $1/\Omega$.

{\em Low Temperature Limit}, $\varepsilon \gg 1$, $\Omega\gg 1$:
In this limit, to leading order in $1/\varepsilon$ and $1/\Omega$
Eq.\ (\ref{eq:integral}) becomes:
\begin{equation}
I(\varepsilon) = {1\over 2}
\log \left(1+ {\varepsilon^2\over \Omega^2}\right) +
{\pi^2\over 6} {\Omega^2-\varepsilon^2\over (\Omega^2+\varepsilon^2)^2}
+ {\pi^2\over 3\Omega^2} \label{eq:IlowT}
\end{equation}
The first term in (\ref{eq:IlowT}) is 
the zero temperature result which correctly incorporates the
above small $\varepsilon$ behavior.
The remaining two terms are of order $T^2$, as in usual FLs.
If the temperature is increased  while 
$\varepsilon \gg 1$ is maintained,
then the result depends on $\Omega$  in a nontrivial way: 
a quick check reveals that in the $\Omega\ll 1$ 
limit $I\approx \pi/\Omega$, as in the zero frequency case.

If one seeks a universal function for $I$, it is 
not difficult
to see that an appropriate form, for all relevant values 
of $\Omega$ and $\epsilon$, is:
\begin{equation}
I(\varepsilon) = {1\over 2}
\log \left(1+ {\varepsilon^2\over \Omega^2}\right)
+ {\pi^2\over 2\Omega(\Omega+\pi)} \label{eq:Igeneral}
\end{equation}
This function displays the correct limiting 
behavior for both low and high frequencies and temperatures.
 As a function of $\Omega$ 
it has a crossover from $1/\Omega$ to $1/\Omega^2$
behavior, reflecting the fact
that as a function of temperature $I$ displays a
crossover from $(\pi T)^2/\omega^2_{kk^\prime}$ at low temperatures to
$\pi T/\omega_{kk^\prime}$ at high temperatures (see Eq.\ (\ref{eq:def_eps})).

The result (\ref{eq:Igeneral}) leads to the correct
behavior of the scattering rate 
 in the case of typical
metallic conductors, for which the relevant energy
scale $\omega_{kk^\prime}$ is of order Fermi energy, 
$E_f$, $\omega_{k,k^\prime}$ is
weakly momentum dependent and $\Omega\sim E_f/T\gg 1$ for 
all $k$ and $k^\prime$.
On using Eqs.\  (\ref{eq:IlowT}) and (\ref{eq:Igeneral}),
one trivially recovers the FL result:
\begin{equation}
{1\over  \tau}  \propto  {\omega^2 + (\pi T)^2
\over E_f^2} 
\end{equation}

It is important to realize that 
the form of the integral $I$ in Eq.\ (\ref{eq:Igeneral})
is such that one can treat the dependence on 
frequency and on temperature almost independently.
When $\varepsilon$ and 
$\Omega$ are of comparable size, the second term in
Eq.\ (\ref{eq:IlowT})
is small compared to the third, and finds a result,
which, although differing somewhat from Eq.\ (\ref{eq:Igeneral}),
\begin{equation}
I(\varepsilon) = {1\over 2} \log \left(1+{\varepsilon^2\over \Omega^2}\right)
+{\pi^2\over 3\Omega(\Omega + \pi/3)}
\label{eq:pi3},
\end{equation}
turns out to be correct for a wide variety of
experimentally relevant parameters. 
For Eq.\ (\ref{eq:pi3})
the crossover from $\pi/\Omega$ behavior at $\Omega\ll 1$ to
$\pi^2/\Omega^2$ behavior at $\Omega\gg 1$ occurs
at considerably lower values of $\Omega$ than is the case with
Eq.\ (\ref{eq:Igeneral}).

For the case at hand, for which $\omega_{kk^\prime}\leq kT$, we follow
Refs.\  \onlinecite{sp-long} and \onlinecite{bps}
and carry out the integration over $k^\prime$ in Eq.\
(\ref{eq:tauI}) by making the approximation 
\begin{equation}
\omega_{kk^\prime} \approx \omega_{sf}\left[ 
1 + \xi^2 (\Delta k^\prime)^2 +
\xi^2(\Delta k)^2\right],
\end{equation}
where $\Delta k^\prime$ and $\Delta k$ measure the
displacements of $k$ and $k^\prime$ 
from the corresponding adjoint hot spots along the FS.
Then, for doping levels such that one has
a large FS,  to a high degree of
numerical accuracy, the integral over 
$\Delta k^\prime$ can be extended to infinity, and
for a general point near the FS the scattering rate is given by:
\begin{equation}
{1\over \tau_k(\omega)} = {\alpha g^2 \xi\sqrt{\omega_{sf}}\over 2 \pi v_f}
\left( {F^\epsilon_k} + {F^T_k}\right)\label{eq:fe_ft}
\end{equation}
where
\begin{equation}
{F^T_k} = {\pi T\over 2} \left[
{1\over \sqrt{\omega_{sf}(1 + \xi^2 (\Delta k)^2)}} - 
{1\over \sqrt{\pi T + \omega_{sf} (1 +\xi^2 (\Delta k)^2)}}\right]
\end{equation}
and
\begin{equation}
{F^\epsilon_k} = 
\sqrt{2} \sqrt{\sqrt{\omega_{sf}^2 [1+ \xi^2 (\Delta k)^2]^2 + 
\omega^2} + \omega_{sf} (1+ \xi^2 (\Delta k)^2)} - 
2 \sqrt{\omega_{sf} (1+ \xi^2 (\Delta k)^2)}.
\label{eq:f_e_k}
\end{equation}
The behavior of $F^T_k$, which determines the low frequency, $\omega$ 
independent limit of 
$1/\tau_k(\omega)$, has been discussed in some detail in
Ref.\ \onlinecite{sp-long};
we turn therefore to the behavior of $F^\epsilon_k$ in various limiting 
cases. 

{\em Finite (but low) frequencies}, $\omega\ll \omega_{sf}$: 
As indicated by the 
$\varepsilon\rightarrow 0$ behavior
of integral $I(k,k^\prime,\varepsilon)$ above,  the low frequency
correction to the (temperature induced) 
scattering rate is always proportional to 
$\varepsilon^2\propto \omega^2$, as in ordinary FLs. However, unlike 
Landau Fermi liquids, from Eqs.\ (\ref{eq:f_e_k}) one finds that the  
coefficient of proportionality, $\alpha_k$, is strongly momentum dependent:
\begin{equation}
1/\tau_k(T,\omega) - 1/\tau_k(T,\omega=0) \approx {\alpha 
g^2 \over 2\pi v_f} 
{\sqrt{\omega_{sf}}\xi \over 4\omega_{sf}^{3/2}
(1+\xi^2(\Delta k)^2)^{3/2}}\,\omega^2\equiv \alpha_k\omega^2
. \label{eq:eps_exp}
\end{equation}
For hot quasiparticles with 
 $\Delta k\xi\ll 1$, the characteristic energy scale which determines
this FL-like behavior is small, being $\sim \omega_{sf}$, while
for cold quasiparticles, with
$\Delta k\xi\gg 1$, it is much larger
$\sim \omega_{sf}\xi^2(\Delta k)^2$. Hence, in the cold regions
$1/\tau$ is proportional to $\omega^2$ up to rather high frequencies
of order $\omega_{sf}\xi^2\sim 100$ meV.
This result has  important consequences: 
although the range of frequencies 
where the FL-like scattering rate
persists is very small in the hot regions, 
one can use FL theory to calculate the effective mass, 
$m^*=m(1-\partial\Sigma^\prime/\partial\omega)$,
which may be large, but is always finite. For 
cold quasiparticles one finds a comparatively modest
effective mass enhancement at all temperatures of interest.

{\em Intermediate frequencies}, 
$\omega_{sf}\ll\omega\ll\omega_{sf}\xi^2$: It follows from Eq.\ 
(\ref{eq:f_e_k}) that the  hot 
quasiparticles possess a lifetime:
\begin{equation}
{1\over \tau_k} \approx {1\over \tau_k(\omega=0)} + {\alpha g^2\over 2 \pi
v_f} \sqrt{2\omega\,\omega_{sf}\xi^2}
\label{eq:q_k_p_k}
\end{equation} 
while for cold quasiparticles, as long as 
$\omega\ll\omega_{sf}\xi^2 (\Delta 
k)^2$, Eq.\ (\ref{eq:eps_exp}) still holds. 

{\em Intermediate frequencies}, $\omega\sim\omega_{sf}\xi^2(\Delta k)^2$,
$\Delta k\leq 1$:
From Eq.\ (\ref{eq:f_e_k}) it is also
relatively easy to obtain a simple expression for the cold quasiparticle 
lifetime 
at intermediate values of frequency and $\Delta k$, such that $\omega\sim 
\omega_{sf}\xi^2(\Delta k)^2$:
\begin{equation}
{1\over \tau_k} - {1\over \tau_k(\omega=0)} \approx {\alpha g^2\over 2\pi
v_f} {\eta\omega\over \Delta k} \label{eq:interm}
\end{equation} 
where $\eta$ is a numerical coefficient of order 0.7.
On combining Eqs.\ (\ref{eq:eps_exp}) and (\ref{eq:interm})
and making use of our previous results for $1/\tau_k(\omega=0)$,
we see that, to a good degree of accuracy, we  can write:
\begin{equation}
{1\over \tau_k(\omega,T)} \approx {\alpha g^2 \over 4 v_f\Delta k}
\left({(\pi T)^2\over \pi T  + \pi T_0(k)} + {\omega^2\over \omega +
E_0(k)}\right) 
\label{eq:tau_approx}
\end{equation} 
where 
\begin{equation}
\pi T_0(k)\sim E_0(k) \sim 2\omega_{sf}\xi^2(\Delta k)^2
\end{equation}
The quantities $\pi T_0$ and $E_0$ are not identical,
but are nevertheless of comparable magnitude.
Hence we see, as might expected for any interacting FL, 
that the same energy scale $E_0(k)$ (up to 
a multiplicative constant of order $\pi$)
 determines the crossover from  FL behavior at low 
$\omega$ and $T$ to a non-Fermi liquid linear in
$\omega$ and $T$ regime. The crossover frequency increases somewhat 
with increasing temperature if, as is the case in overdoped materials,
$\omega_{sf}$ no longer plays a 
negligible role in determining $1/\tau_k(\omega)$. In this case $E_0$ is 
large and $1/\tau$ is only weakly frequency dependent.

As in the case of the d.c.\ resistivity, one expects the dominant 
scattering rates for the optical conductivity 
to be those of the relatively well defined quasiparticles 
in the cold regions of the BZ; the corresponding values of $\Delta k$ lie 
within a range  $\Delta k\sim 1-2$ for an 
overdoped material.\cite{sp-long}  For an underdoped 
material $\Delta k$ is much smaller, 
$\Delta k< 0.5$, since the FS is not only close to the magnetic 
BZ boundary, but there may be strong incommensuration
effects. In addition,  while in the 
overdoped materials $\omega_{sf}\xi^2=const$ as required by the $z=2$ 
scaling, in  underdoped materials one encounters the 
$z=1$ scaling relationship between the characteristic energy and 
momenta, where $\omega_{sf}\xi=const$ and $\xi \sim 1/T$.

A key feature of the NAFL model is
that the relevant scale for the
frequency and temperature variation of the relaxation rates,
$E_0=\omega_{sf}\xi^2(\Delta k)_{max}^2/\pi$ is considerably 
smaller than the fermionic bandwidth. Hence at practically
any frequency of interest (say, $\omega>10$meV)
and in the temperature regime where the resistivity is linear in
$T$ one finds from Eq.\ (\ref{eq:fe_ft})
that the {\em momentum average} scattering rate as a function of
frequency and temperature takes the MFL form:
\begin{equation}
\left<{1\over \tau}\right>_k \approx {\alpha g^2 \over 4\pi v_f} 
(\omega + \pi T).\label{eq:nafl_mfl}
\end{equation}
Thus, the NAFL approach leads to the major result of  MFL 
theory\cite{varma} without resorting to a model in which
(see Eq.\ (\ref{eq:varma})) one encounters an infinite effective mass 
of the quasiparticles; indeed, as noted above, in the NAFL model 
the effective mass of the cold quasiparticles
is always of the order of unity.

Eq.\ (\ref{eq:tau_approx}) implies a parallel between the $T$ 
and $\omega$
dependence of the scattering rates: 
since the same characteristic frequencies 
appear in both cases\cite{sp-long}
 the behavior of $1/\tau$ as a function of $\omega$ or $T$
is approximately the same. 
We verify this in Figure \ref{fig:scatt1} where we plot
the zero temperature
scattering rate as a function of $\omega$ at fixed large $\xi$ and
$\omega_{sf}\rightarrow 0$ for several values of $\Delta k$ near the FS.
The top panel shows quasiparticle scattering rates
when the FS is  large (Eq.\ 
(\ref{eq:q_k_p_k})), while the bottom panel displays our result
for a small effective
FS with 
a maximum distance from a singular point along the FS of $\Delta 
k_{max}=0.3$. 
In the latter case the integral over $k^\prime$ in Eq.\ (\ref{eq:tauI})
was done numerically. The corresponding scattering rates 
as a function of temperature at zero 
frequency are given in the insets of the panels.
The parallel between the zero-frequency temperature and 
zero-temperature frequency dependencies of the scattering rate is
quite obvious. 
Clearly, the rate is highly anisotropic as a function of momentum
and decreases 
dramatically as one departs from the singular point, $\Delta k=0$, where it 
is the largest and most anomalous, $1/\tau\sim \sqrt{\omega}$. On the 
other hand we notice that the influence of the FS size is felt mostly 
in the hot regions while  cold quasiparticle lifetimes are comparatively
insensitive to the size of the FS.

We consider next the influence of changes in the spin fluctuations 
spectrum on the quasiparticle scattering rate. We begin with
systems and temperatures which display 
$z=2$ scaling, where $\omega_{sf}\xi^2$ does not depend on
temperature. As explained in Sec.\ \ref{sec:background}, 
this scaling law is observed in NMR measurements on overdoped systems or 
underdoped systems at temperatures above $T_{cr}$. 
In Figure \ref{fig:hot_cold_z2}a we show  
the frequency dependence of the hot quasiparticle scattering rate for
several different temperatures.  In order to make the connection with 
experiment, in this, and many subsequent figures 
we plot the inverse mean free path  in the units of lattice 
spacing, by assuming a constant Fermi velocity, $v_f=0.25$eV.
As in previous work,\cite{sp-long}
we have assumed realistic spin-fluctuation
parameters\cite{NMR-refs} appropriate for YBa$_2$Cu$_3$O$_7$,
$\omega_{sf}=[6+0.06\,T\,(K)]$ meV
and $\omega_{sf}\xi^2=60$ meV.
 We see that the shape of the curve 
is somewhat less anomalous
than that shown in Fig.\ \ref{fig:scatt1}.
Since the onset of the non-FL
frequency dependence is at  $\omega\sim \omega_{sf}$ for a hot spot, 
the observed behavior is a
 natural consequence of using a finite value of 
 $\omega_{sf}$ and  a correspondingly modest value of $\xi$. 
Since $\omega_{sf}$ increases with temperature
the scattering rate becomes even more FL-like at higher
 temperatures . 
Quite generally, outside of the pseudogap regime,
the behavior of hot quasiparticles becomes
less anomalous as $T$ increases, since the system
is further away from the AF instability, as measured directly by the size
of $1/\xi$ and/or $\omega_{sf}$. Nevertheless, a close inspection
shows that the behavior of $1/\tau(\omega)$ is sublinear, i.e.,
$1/\tau\propto \omega^a$ where $a<1$.

Figure \ref{fig:hot_cold_z2}b shows the comparable result for cold 
quasiparticles: we display the average scattering rate averaged
over values of 
$\Delta k$, such that $0.5/\xi<\Delta k < 2/\xi$, for the same temperatures 
as in Fig.\ \ref{fig:hot_cold_z2}a.  This average scattering rate
is a good measure of the 
effective relaxation rate, $1/\tau_{eff}(\omega)$, seen in optical 
measurements, since the 
behavior of quasiparticles at  intermediate values of $\Delta k$, i.e.,
$(\Delta k)_{max}\sim 0.5-1$, dominates the transport.   
Aside from 
numerical value, which is  considerably reduced
  from that found at the hot spots
(see Fig.\ \ref{fig:hot_cold_z2}c), we see that
the behavior is FL-like up 
to a frequency of order 15 meV, becoming linear with increasing $\omega$.  
There is no sign of sublinear behavior in $1/\tau$.

In Figure \ref{fig:hot_cold_z2}c,  we have plotted
the relaxation times for both hot and cold quasiparticles in 
YBa$_2$Cu$_3$O$_7$ at
$T=150$K, in order to demonstrate the considerable  variation
in quasiparticle behavior as one moves around the Fermi surface:
the frequency dependence of the hot and cold quasiparticle
lifetimes are seen to be both qualitatively and quantitatively
different. Hot quasiparticles possess a  mean 
free path  of  order of a few lattice spacings and are comparatively ill
defined at higher $\omega$, while in the cold regions of the FS  we find
very well defined quasiparticles, with mean free paths of up to ten lattice 
spacings. Note that the mean free path is proportional to $v_f^2$ and hence
 is rather sensitive 
to the choice of this parameter. Our choice of $v_f=0.25$eV is somewhat 
conservative, and the exact value of the mean free path 
is probably somewhat larger
than that displayed in the Figure. 

We note that, due to the FL-like behavior at low frequency, 
the behavior of $1/\tau(\omega)$ for cold quasiparticles 
at different temperatures becomes increasingly more distinct as $T$ 
increases.  This is in sharp contrast with the behavior observed in the 
singular (hot) region, especially when $\xi$ acquires the temperature 
dependence appropriate to the $z=1$ scaling case.  This is shown in Figure 
\ref{fig:hot_cold_z1}a, where we plot our result using spin fluctuation
parameters
appropriate for the  underdoped YBa$_2$Cu$_4$O$_8$ 
material.\cite{BP}  The 
temperature dependence of $1/\tau$ is now completely suppressed at higher 
frequencies and eventually even inverted from the conventional
FL behavior, especially at larger $T$.  By contrast, for the cold 
quasiparticles, as shown in
Fig.\ \ref{fig:hot_cold_z1}b,  we obtain results which are very similar 
to those shown in Fig.\ \ref{fig:hot_cold_z2}b.  In Sec.\ 
\ref{sec:sigma_xx} we will show that these results lead to the observed 
optical properties in HTSs.\cite{timusk-review} In particular
we show that $1/\tau_{eff}$ is linear in $\omega$ regardless of the 
scaling law and hence, provided that the vertex corrections are not large,
transport in
underdoped materials above $T_*$ is  very similar to that in optimally 
doped materials.

\section{Optical Conductivity}
\label{sec:sigma_xx}
\typeout{Opt cond on page \thepage}

We now turn to the optical conductivity.
We begin with the Kubo formula and  the well-known 
expression for the optical conductivity in the absence of the 
vertex corrections:\cite{rickayzen}
\begin{equation}
\sigma_{\mu\nu}(\omega) = {ie^2\over 2\pi^2\omega N_k}
\sum_{k} v_k^\mu v_k^\nu
\int d\omega^\prime d\omega^{\prime\prime}
A_k(\omega^\prime) A_k(\omega^{\prime\prime})
{f(\omega^{\prime})-f(\omega^{\prime\prime})\over \omega-\omega^{\prime}+
\omega^{\prime\prime}-i0_+}
\label{eq:sigmamunu}
\end{equation}
where $N_k$ is the number of points in the BZ and $v_k=\nabla \epsilon_k$,
which in the case of a four-fold symmetric system, where $x$ and $y$ 
directions are equivalent, yields for the real part 
of $\sigma_{xx}(\omega)$:
\begin{equation}
Re \sigma_{xx}(\omega) = {e^2\over 4\pi N_k}
\sum_{k} v_k^2
\int d\omega^\prime 
A_k(\omega^\prime) A_k(\omega^{\prime}+\omega)
{f(\omega^{\prime})-f(\omega^{\prime}+\omega)\over \omega}.
\label{eq:resigmaxx}
\end{equation}
One can, of course, obtain  $Im\, \sigma$ using the Kramers-Kronig
relations.
In Eqs.\ (\ref{eq:sigmamunu}) and (\ref{eq:resigmaxx})
$A_k(\omega)$ is the single particle spectral function,
\begin{equation}
A_k(\omega) = {\vert Im\Sigma_k(\omega)\vert \over 
[\omega-\epsilon_k-Re\Sigma_k(\omega)]^2 + [Im\Sigma_k(\omega)]^2}
\label{eq:spectral}
\end{equation}
and $\Sigma_k(\omega)$ is the single particle self-energy.

In general, Eq.\ (\ref{eq:resigmaxx}) is hard to solve 
analytically. However, depending on 
the properties of the self-energy,
various approximations can be used. For example,
if the self energy is only weakly momentum and frequency
dependent, one can use the relaxation time approximation:
\cite{rickayzen}
\begin{equation}
\sigma_{xx}(\omega) = {e^2\over N}
\sum_{k} v_k^2 {1\over 1/\tau_k(\epsilon_k) - i\omega} \left({\partial 
f\over\partial \epsilon}\right)
\label{eq:sigmarelax}
\end{equation}
which is the result one obtains using
the Boltzmann transport theory. 
It has been argued that qualitatively the optical properties of
cuprates can be captured within this formalism, although, 
assuming a weak dependence of $\Sigma(k,\omega)$ on $k$ and $\omega$
clearly represents a considerable oversimplification.

An interesting result can be obtained from 
Eq.\ (\ref{eq:sigmarelax}). On performing 
the appropriate integrals, at low $T$,  we find that
\begin{mathletters}
\label{eq:sigma_xx}
\begin{equation}
Re \sigma_{xx} = e^2{C T\over \omega^2} \log\left[1+ \left(K_{max} 
\omega\over
CT\right)^2\right]
\end{equation}
and
\begin{equation}
Im \sigma_{xx} = e^2 \left( {K_{max}\over \omega} + CT 
\tan^{-1}{CT\over K_{max}\omega} - CT {\pi\over 2}\right)
\end{equation}
\end{mathletters}
where $C=4v_f/\alpha g^2$.
We see that, in contrast to ordinary FLs,
in NAFLs the real part of $\sigma_{xx}$ at high frequency
depends only weakly on temperature.
This weak
temperature dependence of $\sigma_{xx}(\omega)$ at high frequency is a 
generic feature of any system with a low energy soft mode, be it a 
NAFL, as above, or a system near a charge density wave 
instability.
Note that Eq.\ 
(\ref{eq:sigma_xx}) reduces to the appropriate FL results in the 
$\omega\rightarrow 0$ limit.

A generalization of the relaxation time approximation,
the memory function formalism,\cite{shulga} is appropriate
for a self energy which, in the vicinity of the FS,
does not depend on momentum, as is the case
when the dominant scattering mechanism
is due to phonons, as in ordinary 
metallic superconductors; at sufficiently high
frequency, it leads to a Drude-like conductivity:
\begin{equation}
\sigma_{xx}(\omega) = {ne^2\over 1/\tau_{eff}(\omega) 
- i\omega \lambda(\omega)}
\label{eq:memory}
\end{equation}
with a frequency dependent effective scattering rate, and a 
frequency dependent mass $m_*$, 
$m_*=m_e(1+\lambda(\omega))$.
This expression is not directly applicable
for the case of highly anisotropic scattering rates such as those studied 
here. However, we show below 
that in NAFLs, at sufficiently high frequencies,
one can obtain an expression of this form, in which
$1/\tau_{eff}$ represents an average scattering
rate as a function both of momentum and frequency.

We proceed in the following way:
We first recall that in the NAFL model 
the imaginary part of the self energy 
depends only weakly on the component of the wavevector $k$
perpendicular to the FS; 
the dominant momentum dependence comes from the component which lies
in a direction parallel to 
the FS (see  Eq.\ (\ref{eq:f_e_k}) and the discussion which follows).
This is clearly an 
approximation, but not an unreasonable one, since 
only quasiparticles in a narrow region (of order $T$) around the FS are 
involved in  the scattering
(see Eq.\ (\ref{eq:resigmaxx})).
In this case one can perform the change of variables (\ref{eq:varchange})
and integrate over $\epsilon^\prime$ explicitly. If the integral can be 
extended to infinity, the result can be expressed in the quite compact form:
\begin{equation}
Re\sigma_{xx}(\omega) = {e^2\over 8\pi^2}
\int_{FS} {dk\over \vert v_f\vert }
\int d\omega^\prime 
{f(\omega^{\prime})-f(\omega^{\prime}+\omega)\over \omega}
Im\left( {v_f^2\over \omega - 
\Sigma_k(\omega+\omega^\prime)+\Sigma_k^*(\omega^\prime)}\right)
\label{eq:resigmakk}
\end{equation}
In the limit $\omega\tau\gg 1$ we can expand this expression and compare 
our result to the Drude-formula in the same limit. We find:
\begin{equation}
{1\over\tau_{eff}(\omega)} = \int_{FS} {dk\over 2} \int d\omega^\prime
{f(\omega^{\prime})-f(\omega^{\prime}+\omega)\over \omega}
\left({1\over \tau_k(\omega+\omega^\prime)} + {1\over 
\tau_k(\omega^\prime)}\right)
\end{equation}
In the limit $\omega\gg T$ this yields
\begin{equation}
{1\over \tau_{eff}(\omega)} = \left< {1\over \tau_k}\right>_{\omega,k}
\end{equation}
where the brackets indicate the appropriate average. We have already 
obtained the momentum averaged scattering rate, Eq.\ 
(\ref{eq:tau_approx}), which, at high frequencies, 
took the MFL form, Eq.\ (\ref{eq:nafl_mfl}).
These expressions show that the average (over momentum) of $1/\tau_k$ is a 
simple power law as a function of frequency, being quadratic and linear
at low and high $\omega$ respectively. 
As a result, averaging over frequency will not qualitatively
change the frequency dependence 
of the momentum averaged scattering rate,
$<1/\tau_k>_k$, since an average of a power-law function is that same 
function. Therefore, aside from numerical factors (of order 2) 
and a somewhat different crossover scale $E_0$, the momentum averaged
scattering rate $<1/\tau_k>_k$ is a reasonable approximation
for $1/\tau_{eff}(\omega)$ in the  limit, $\omega\tau_{eff}\gg 1$.

Several important conclusions can be drawn from this result.  
First, only at high frequencies should the Drude formalism be 
used to analyze 
experiments. 
But, since at 
sufficiently high frequencies, in general, 
the quasiparticle picture breaks down, the results we obtain are likely 
only qualitative.
One can show that Drude-like behavior is maintained as one goes to 
lower frequencies, but the corresponding scattering rate is no longer
simply related to the quasiparticle self-energy, and 
will not be frequency independent. 
Moreover, quite importantly, in the limit $1/\omega\tau\ll 1$ any feature 
seen in experiments, such as the weak temperature dependence of 
$1/\tau_{eff}(\omega)$ at higher $T$ and $\omega$, or even the pseudogap 
seen in underdoped materials, must stem from
the quasiparticle lifetimes analyzed in the previous section, since it 
is their average which is probed.

\section{Comparison with Experiment}
\label{sec:comparison}
\typeout{Comparison on page \thepage}

We now compare our results with experiment, in order to 
test the applicability of NAFL theory when one uses
realistic parameters for both the effective interaction
and band parameters and carries out the relevant integrals numerically.
Since the optical conductivity as a function of frequency frequently
appears featureless, it is customary to use the memory function 
formalism\cite{timusk-review}
to extract the effective transport scattering rates 
from  experiments. Where this has been done,
the analysis of the previous section
allows us to compare directly our (averaged) scattering times with
experiment. 

For systems for which the scattering rates have not been 
determined we compare our calculated values of $\sigma_{xx}(\omega)$
directly to  experiment. In so doing, we make use of our numerical results
for the optical conductivity 
obtained by substituting the self energy 
(\ref{eq:scatkq}) into Eq.\ (\ref{eq:resigmaxx}), with and without
including the inclusion of 
the vertex $\gamma_{k,k^\prime}=\cos\theta$ for $1/\tau_k$
where $\theta$ is the angle between $k$ and $k^\prime$. Strictly speaking, 
this form of the
vertex correction is only applicable if the effective interaction is 
weakly frequency dependent, since only then 
is the current vertex  proportional 
to the current,  and hence
additive to the self-energy. In NAFLs this is 
 a somewhat 
 questionable approximation, but as long as $qv_f\ll 1$, while 
$V_{eff}$ is highly peaked near $Q$ and the FS is large, 
the magnitude of the 
vertex correction turns out to be 
comparatively small, leading  to corrections 
of order 10\% for systems near optimal doping. However, since 
it is straightforward to include it numerically, we 
retain the function $\gamma_{k,k^\prime}$ in the subsequent 
numerical results.

We begin with overdoped materials.  While it is extremely important to 
compare the calculated optical conductivity with experiment
only for the samples which are 
used to determine the spin fluctuation parameters, NMR and optical
experimental results on the same sample are 
often not available. Thus for the first sample we consider, single layer 
Tl 2201, the samples used by Puchkov et al,\cite{timusk-review} for  
optical experiments had a $T_c=15$K, while the best available NMR 
experiments on a comparable sample are those of Itoh et al,\cite{itoh} 
on a Tl 2201 sample which had a $T_c=23$K. For overdoped systems, the 
various parameters ($\alpha$, $\omega_{sf}$, $\xi$) which determine the 
spin fluctuation spectrum do not vary appreciately with the doping level; 
hence our use of parameters chosen to fit the NMR experiments of Itoh et al 
should not produce an appreciable error. In choosing band structure 
parameters for our calculation, we assumed the band structure for the Tl 
system is close to that found in BSCCO family. Our calculated results, 
are compared with experiment in  
Fig.\ \ref{fig:tl_tau}.
We present our results only for $T>100$K, since for $T\ll 100$K the stability
of our numerical solution becomes questionable.
Clearly the fit is quite good; both theory and experiment show
that the system crosses over from the quadratic to linear in frequency 
behavior at a frequency $\pi T_0$, which is of order $\omega_{sf} \xi^2(\Delta 
k)_{max}^2/2=40$meV.
 Note that the crossover behavior changes very little with
temperature,  as predicted by  Eq.\ (\ref{eq:tau_approx}) for the case of 
the $z=2$ scaling present in all (sufficiently) overdoped materials.
We further note that only two fitting parameters appear in our
calculation: the coupling 
constant, $g$,  and the zero frequency 
offset (which is very small, $1/\tau(\omega=0)\sim 30$ cm$^{-1}$ and is
likely generated by impurities). If the resistivity 
were measured for this material  both fitting parameters would be
fixed by experiment.

In Fig.\ \ref{fig:gins} we compare our  calculated optical conductivity
for optimally doped YBa$_2$Cu$_3$O$_7$ with the experimental results 
of Ref.\ \onlinecite{gins-optical}. We have assumed, as in previous work, 
that  $\omega_{sf}=[6+0.06\, T\,(K)]$meV, $\omega_{sf}\xi^2=60$meV. Strictly 
speaking this scaling law,
applicable only for temperatures above $T_{cr}\sim 140$K, is  only 
approximately correct at lower temperatures.
At high temperature our calculated 
results agree quite well with the experiment, both qualitatively and 
quantitatively.  At low $T$  qualitative agreement is found; 
 the calculated $\sigma_{xx}(\omega)$ at $T=100$K
crosses the calculated $\sigma_{xx}(\omega)$ at  $T=300$K  at roughly 
the same frequency as that seen experimentally, 
although at high $\omega$  the calculated values of $\sigma_{xx}$
are considerably lower than those seen  experimentally.  
We  attribute this discrepancy, in part, to the possible 
presence of disorder in the experimental samples: for example,
we found it 
necessary to use a higher value of the coupling constant 
$g$ ($g\approx 0.6$eV) than that used in Ref.\ \onlinecite{sp-long}
($g\approx 0.5$eV) to fit the 
resistivity. A small amount of, e.g., impurity scattering
can bring about significant changes in transport and optical behavior
at relatively low temperatures, while producing a negligible effect
at higher $T$.  

An interesting feature of the YBa$_2$C$_x$O$_y$ family is
the a-b plane anisotropy, which is  seen in untwinned single crystals
near optimal doping. 
[We have earlier provided a qualitative 
explanation for the anisotropy in the
resistivity of these materials\cite{sp-long} and it is instructive
to consider its optical counterpart.]
In Fig.\ \ref{fig:1248} we show the optical conductivity for 
a model system
which displays strong $a-b$ planar anisotropy, comparable to
that observed in YBa$_2$Cu$_4$O$_8$.
We follow Ref.\ \onlinecite{sp-long}
and assume that the main influence
of the CuO chains is to modify the electronic properties 
of the a-b plane and so model the observed anisotropy
by introducing  anisotropic
hopping matrix elements 
along the two different crystallographic directions, with $t_a=0.55 t_b$. 
This value of $t_a$ is somewhat low, but is  not unreasonable
for a qualitative analysis  for pedagogical purposes.
The input parameters ($\omega_{sf} = [2 + 0.02\, T\,(K)]$ meV
and $\omega_{sf} \xi^2=60$ meV) are the same as 
in our earlier work.\cite{sp-long}
As can be seen from the figure, the strong anisotropy we found for 
the d.c.\ resistivity has its analog  in the case of 
$\sigma_{xx}(\omega)$, as
might have been expected;  the strongly conducting $a$
direction displays 
a distinctly
different $\omega$ dependence at higher frequencies. This 
behavior can be
understood by recalling that when 
the current runs along the $a$  direction, 
the distribution of hot spots is such as to
yield scattering rates which are mostly FL like, 
and hence one should expect to see a large conductivity even at large 
$\omega$. On the other hand, 
for current running along the $b$ direction the dominant contribution 
to the conductivity comes from
quasiparticles which possess  strongly 
anomalous scattering rates; this leads to 
a much stronger frequency dependence of
$Re \sigma_{bb}(\omega)$, of the kind observed
in  many underdoped samples.  

We consider next the underdoped materials.  Above 
$T_*$, we expect the behavior of $1/\tau_{eff}$ to take  the MFL 
form, Eq.\ (\ref{eq:nafl_mfl}). Indeed, Eq.\ (\ref{eq:nafl_mfl}), obtained 
for a moderately
large FS, does not depend on the spin fluctuation 
parameters, $\omega_{sf}$ 
and $\xi$, and hence independent  
of the  scaling law relating $\omega_{sf}$ and $\xi$;
$1/\tau_{eff}$ is linear in both $T$ and 
$\omega$ both above and below $T_{cr}$, as long as 
$T>T_*$.
On the other hand, it is now widely believed that in 
heavily underdoped materials, below $T_*$,
the hot
quasiparticles, which are
close to the  magnetic BZ boundary become effectively gapped (the 
pseudogap), with much of the quasiparticle weight transferred from  low
to high frequencies; the quasiparticle band dispersion thus
acquires a form similar to that encountered with
a preformed spin-density-wave (SDW) 
state.\cite{chub-morr}  Hence, at low frequencies, 
hot quasiparticles barely contribute to transport, a conclusion 
which is consistent with 
the superfluid density measurements at low temperatures, which 
suggest that only 
a fraction of the doped holes become part of 
the superconducting condensate.\cite{MP-pseudo,tanner}
Although quasiparticles are not well defined
in the hot region of the BZ, since their quasiparticle residua are rather small,
 they still contribute to transport
up to frequencies of order $T_*$. Above this scale
their scattering is
so strong that $Im \Sigma$ is essentially temperature and 
frequency independent 
(see the result for $\Delta k=0$ in Fig.\ \ref{fig:scatt1}). 
On the other hand, the cold quasiparticles
are still well defined, albeit in a somewhat narrower region, centered around
${\bf k}=(\pi/2,\pi/2)$, and 
 symmetry related points in the Brillouin zone. 
However, since so much of the 
FS is gapped,
the phase space available for the scattering of cold 
quasiparticles is considerably reduced.
Thus for small frequency, $\omega<T_*$,
one  expects $1/\tau_{eff}\propto \omega^2$, just as one finds $1/\tau\sim T^2$ at 
zero frequency and temperatures well below $T_*$ in underdoped materials.
In other words, there is a significant temperature 
variation of $1/\tau_{eff}$ up to this 
frequency. 
As the hot quasiparticles cease to contribute to transport
above this energy scale, which is valid
as long as the correlation 
length is large (that is, for $\xi\geq 2$, and $T<T_{cr}$),
the slope of $1/\tau_{eff}$ is reduced 
and it acquires its usual linear in $\omega$ dependence.
In terms of the NAFL model, a close study shows that 
the effective rate $1/\tau_{eff}(\omega)$ increases 
up to frequencies of order 
$T_*\sim \omega_{sf}\xi^2(\Delta k)_{max}^2/2$, where $\Delta k_{max}$ is  much 
smaller ($\sim 0.1$-0.5) than is found in  optimally doped materials.
This phenomenon is the optical analog of the resistivity
which is a considerably stronger function of temperature below $T_*$
than above it.
It is noteworthy that at sufficiently high temperature,
above $T_{cr}$, much of the quasiparticle weight is again transferred to the
low frequencies and the hot quasiparticle again start to contribute to transport,
but  the slope of $1/\tau_{eff}$ does not change, since it 
does not depend on the scaling law between the spin fluctuation parameters.

We note that the fact that the contribution of the hot regions is
strongly reflected in experiment
in the underdoped materials suggests that large parts of the FS are in fact
hot. Moreover, at higher frequency and low temperatures
the slope of $1/\tau_{eff}$  is the same as 
that found at high $T$ in systems
where there is no pseudogap present. If large parts of the FS exhibit 
highly anomalous behavior, then one should expect that at
relatively low frequencies $1/\tau_{eff}$ would begin to saturate.
Indeed, we have found that the 
MFL form of $1/\tau_{eff}$, Eq.\ (\ref{eq:nafl_mfl}),
is obtained only for a large FS, while for a smaller FS one would expect 
$1/\tau_{eff}$ to display sublinear behavior as a function of frequency.
This suggests that the linear in $\omega$ behavior of 
$1/\tau_{eff}$ above $T_*$ reflects  scattering 
processes involving momentum transfers  far from 
${\bf Q}$, which supplement
the magnetic scattering we have considered.
If one is to maintain the
linear behavior of $1/\tau_{eff}$ for $\pi T$ and $\omega$ larger 
than $\omega_{sf}\xi^2$, 
the dominant scattering mechanism must crossover from processes involving 
small to those involving larger 
characteristic frequencies. For example, in 
Ref.\ \onlinecite{sp-long} we introduced  FL scattering in order to 
avoid resistivity saturation at higher temperatures in 214 systems and 
obtained thereby
quite reasonable agreement with experiment near optimal doping 
level.
In this respect, 
optimally doped materials are merely those which have the smoothest 
transition from  low energy to high energy scales, in excellent
agreement with the experimental findings of Takagi et al.\cite{batlogg}

The additional scattering processes introduced to 
maintain linear in frequency $1/\tau_{eff}$ above the pseudogap scale,
do not qualitatively change the behavior of the effective relaxation rate 
at low frequencies which is still governed by the strong magnetic scattering.
We present in Fig.\ \ref{fig:tau_fl} the result of a numerical calculation 
of $1/\tau_{eff}$, where we have used $\Delta k_{max}=0.4$,  
taken $\omega_{sf}$ and $\xi$
to be the same as in Fig.\ \ref{fig:hot_cold_z1}a, and have incorporated
in our numerical work, based on Eqs.\ (\ref{eq:sigmamunu}) and 
(\ref{eq:resigmaxx}),
a  FL-like scattering potential,
$V_{FL}=\chi_0/(1-i\pi\omega/\Gamma)$, where $\Gamma=400$ meV,.
Notice that $1/\tau_{eff}$ loses its temperature dependence
at high frequency, i.e., all curves shown in the figure converge at
a frequency $\sim\omega_{sf}\xi^2(\Delta k_{max})^2/2$. Below
this frequency $1/\tau_{eff}$ displays 
considerable temperature dependence,
even though we have not included the increase of $\omega_{sf}$
at low temperatures and a possible temperature variation in the 
effective size of the FS.
Within the magnetic scenario, these scattering 
processes are closely related to the experimentally measured
energy scales for ${\bf q}\approx {\bf Q}$.
Moreover, since the scale of $\omega_{sf}\xi^2$ is weakly doping
dependent and always of the order of $1/\pi$ times
the Fermi energy
the crossover energy should be 
almost material independent, in agreement 
with NMR experiments which show that
for most materials $\xi(T_{cr})\approx 2$.

\section{Conclusion}
\label{sec:conclusions}

In summary, we have applied both analytical and numerical 
techniques to the study of 
the transport properties of HTSs at optical 
frequencies within the NAFL model. We find that the peaking
of the effective magnetic interaction leads to strong
anisotropy
of the quasiparticle properties in different regions of the BZ, and brings 
about a complex morphology for the optical conductivity and the effective 
scattering rate, $1/\tau_{eff}(\omega)$, which can be extracted from it.
We found that for a wide region of temperature and frequencies
the NAFL model leads directly to the MFL
form for $1/\tau_{eff}$, Eq.\ (\ref{eq:nafl_mfl}), 
a scattering rate which agrees with 
experimental results found in nearly all optical measurements to date.
Overall, in the 
appropriate limits, our results agree qualitatively and many cases 
quantitatively, with the experimental findings.

In this paper we have focused on the optimally doped and overdoped 
materials. In the underdoped case, to obtain results of comparable quality,
one must parametrize in some detail changes in the quasiparticle behavior
brought about by the quasiparticle pseudogap, which we have not yet done.
However, to the extent that the FS undergoes a 
substantial transition as one varies doping from optimal to underdoped
we have shown that the magnetic scenario provides  results in 
qualitative agreement with experiment, even within this simple theory.

Our use of the Born approximation, would seem to work somewhat better than 
one would expect it to
in a system with strong correlations. Ideally,
one should obtain
$\Sigma(k,\omega)$ self-consistently, before substituting it into  Eq.\ 
(\ref{eq:resigmaxx}). However, it may prove easier to calculate
the current-current correlation
function directly.\cite{MPII} The results of  calculations which include both 
the pseudogap and strong coupling effects will be presented 
in forthcoming publications.\cite{joerg-pg,vlad}

\acknowledgements

We are indebted to  G.\ Blumberg, G. Boebinger,
A.\ Chubukov,   D.\ Ginsberg,  
P.\ Monthoux,
T.\ M.\ Rice, J.\ Schmalian, Q.\ Si, C.\ Slichter,
R.\ Stern and T.\ Timusk for stimulating conversations on 
these and related topics.
We should like to thank Tom Timusk for his helpful remarks following a critical reading
of an earlier version of this manuscript.
We thank the National Center for
Supercomputing Applications for a grant of computer time.
This research has been supported in part by NSF through
grants NSF-DMR 89-20538 (MRL at UIUC) and NSF-DMR 91-20000
(STCS).

\begin{figure}
\caption{A model FS in cuprates. For a quasiparticle interaction 
peaked at ${\bf Q}$, the intersection of the magnetic Brilloiun zone 
(shown with the dashed line) with the FS specifies 
the singular points at which the interaction
is maximally effective in low energy scattering processes.}
\label{fig:fs}
\end{figure}

\begin{figure}
\caption{The frequency dependence of the 
quasiparticle scattering rate at $T=0$
in (a) the case of a 
large FS. (b) the case of a small FS, for which $\Delta k_{max}=0.3$.
In both panels the curves correspond to 
(from top to bottom) $\Delta k=0$, 0.1, 0.25, 0.5, 1. The insets show 
the corresponding results  at zero frequency as a function of temperature.}
\label{fig:scatt1}
\end{figure}

\begin{figure}
\caption{The frequency dependence of the 
quasiparticle inverse mean free path (in units of inverse lattice spacing), 
calculated for $\hbar v_f=0.25$ eV,
at (from bottom to top) $T=100$, 200 and 300K. 
The spin fluctuation parameters ($\omega_{sf}=[6+0.06\, T\,(K)]$ meV, 
$\omega_{sf}\xi^2=60$meV)
are appropriate for YBa$_2$Cu$_3$O$_7$.
Panel (a) shows the case with $\Delta k=0$ (hot region); 
panel (b) shows the momentum averaged inverse mean free path for cold 
quasiparticles ($0.5/\xi <\Delta k <2)$; in panel (c) we compare the frequency
dependences of this averaged 
cold quasiparticle mean free path (bottom) with that of
hot quasiparticles (top) at $T=150$K.
Note the separability of the
$T$ and $\omega$ contributions to the scattering rate in panels (a) and (b).}
\label{fig:hot_cold_z2}
\end{figure}

\begin{figure}
\caption{The quasiparticle inverse mean free path
as a function of $\omega$ at (from bottom to top, on the left hand side)
$T=100$, 200 and 300K in an underdoped material. 
The spin fluctuation parameters 
($\omega_{sf}=[3.4 + 0.043\, T\,(K)]$ meV, and $\xi=50meV / \omega_{sf}\approx
0.07 + 0.00086\,T(K)$) are appropriate for the pseudoscaling regime in 
a YBa$_2$Cu$_4$O$_8$ sample. Panels (a) and (b) correspond to 
the hot and momentum averaged cold regions.
Note the unusual temperature dependence of $1/v_f\tau$
at high frequency for the hot quasiparticles.}
\label{fig:hot_cold_z1}
\end{figure}

\begin{figure}
\caption{A comparison of the calculated
scattering rate as a function of $\omega$ in 
Tl 2201 with the experimental results of Puchkov et al.$^1$
The curves correspond (from bottom to top) to 
$T=120$K, 200K and 300K respectively. The 
theoretical curves were obtained using  spin fluctuation parameters
appropriate for this material, $\omega_{sf}=[12+0.003\, T\,(K)]$ 
meV and $\omega_{sf}\xi^2=50$meV. 
The FS size was estimated from the ARPES results on the closely related
BSCCO compounds, for which $\Delta k_{max}\approx 1.4$. The 
coupling constant $g^2$ was adjusted to yield the correct spread of the 
curves at $\omega=0$ and a constant term was added, to take 
into account the likely presence of imperfections in the samples.}
\label{fig:tl_tau}
\end{figure}

\begin{figure}
\caption{A comparison of the calculated optical
conductivity of YBa$_2$Cu$_3$O$_7$ 
with the experimental results of Tanner et
al.$^{20}$ The top curve and the circles correspond to $T=100$K and the bottom 
curve and the squares correspond to $T=300$K. The
parameters are those used earlier,$^9$ $\omega_{sf}\xi^2= 
60$ meV, $\omega_{sf}=[6 + 0.06\, T\,(K)]$ meV, with $t=250$meV and 
$t^\prime=-0.4t$.
Just as was the case for the resistivity, agreement 
with experiment is much better at higher temperatures. 
The value of the coupling constant $g$ used here, 
$g=0.6$eV, is somewhat larger than $g=0.53$eV used in 
Ref.\ 9 
 to fit the resistivity, $\rho_{xx}$. }
\label{fig:gins}
\end{figure}

\begin{figure}
\caption{Schematic behavior of a system with strong a-b plane 
anisotropy, such as YBa$_2$Cu$_4$O$_8$ 
material. The lower (upper)
curve corresponds to $\sigma_{aa}$ ($\sigma_{bb}$), with the 
solid (dashed) line referring to $T=100$K ($T=300$K). 
The input parameters ($\omega_{sf}=[2 +0.02\,T\,(K)]$ meV, $\omega_{sf}\xi^2=60$ meV, 
$t_a = 0.55 t_b$) are the same as 
in Fig.\ 22 of Ref.\ SP. 
Clearly, both the magnitudes and the qualitative behavior
of the two conductivities are different:
$\sigma_{bb}$ is far more FL-like than is   $\sigma_{aa}$.}
\label{fig:1248}
\end{figure}

\begin{figure}
\caption{$1/\tau_{eff}$, obtained for a small FS, $\Delta k_{max}=0.4$,
by numerically solving 
Eqs.\ (\ref{eq:fe_ft}) and (\ref{eq:sigmamunu}), for the same
input parameters as in Fig.\ 4, but adding a  FL scattering
contribution, $V_{FL} = \chi_0/(1-\i\pi\omega/\Gamma)$, with $\Gamma=400$ meV,
to the magnetic interaction, Eq.\ (4).
The curves correspond (bottom to top) to $T=100$, 150, 200, 250K.
At high frequency the averaged quasiparticle 
lifetime is nearly $T$ and $\omega$ independent,
in accord with 
experiment
on the underdoped materials. The stronger temperature dependence at
low frequency stems from the strongly temperature dependent
spin fluctuation spectrum found in
all underdoped materials.}
\label{fig:tau_fl}
\end{figure}

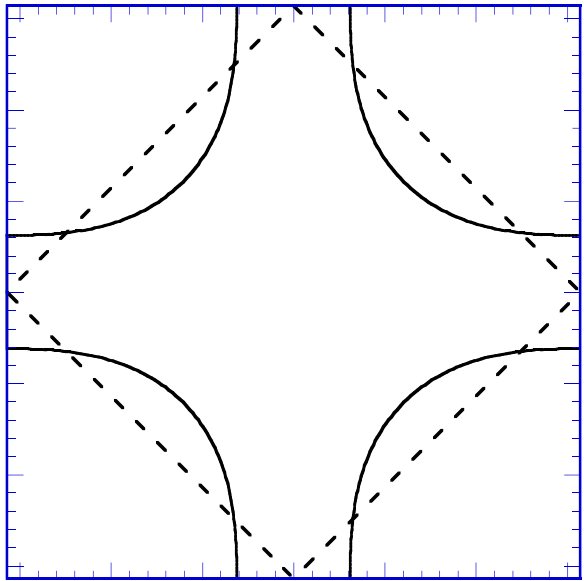
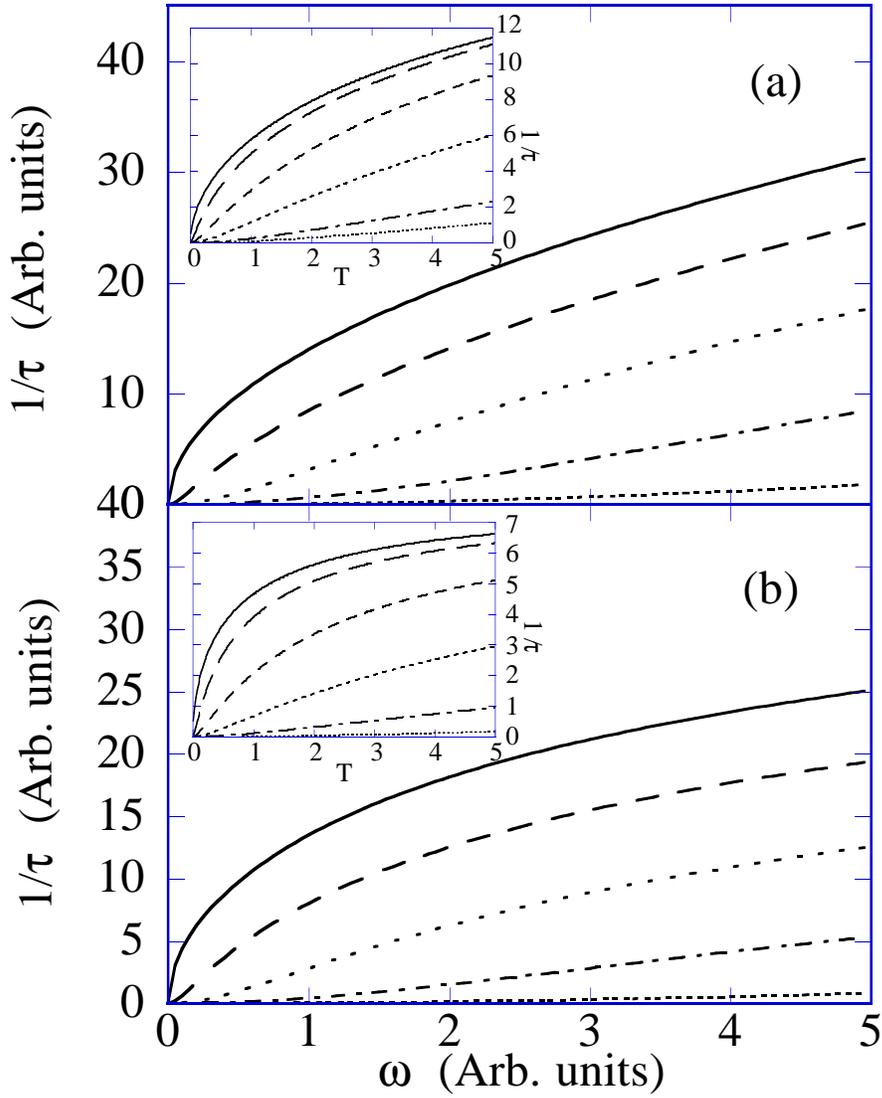
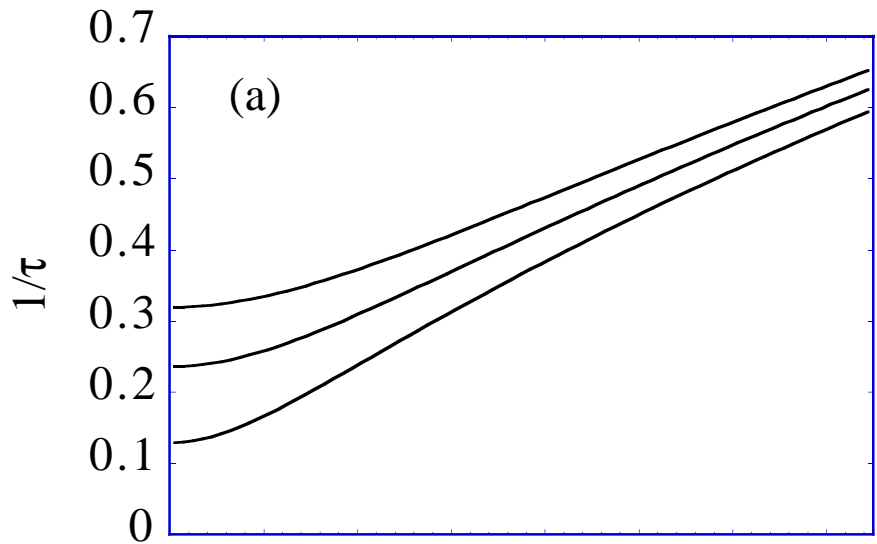
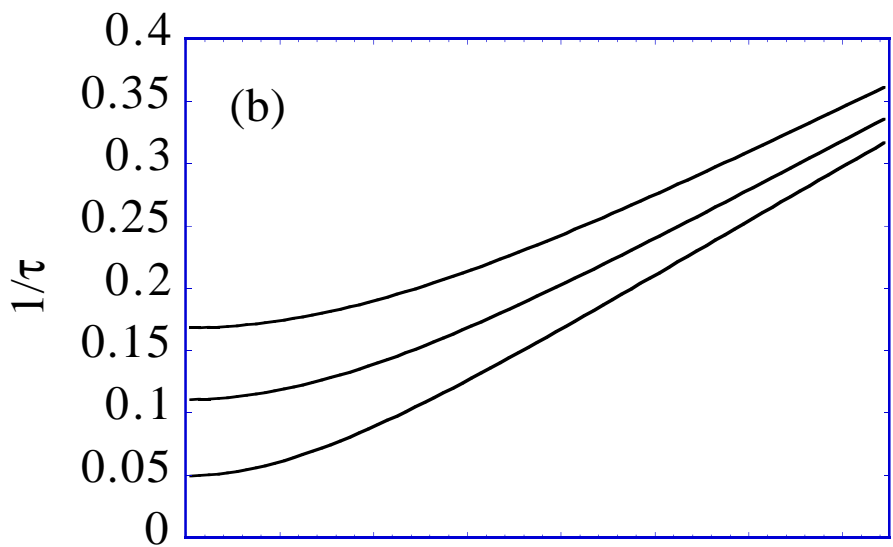
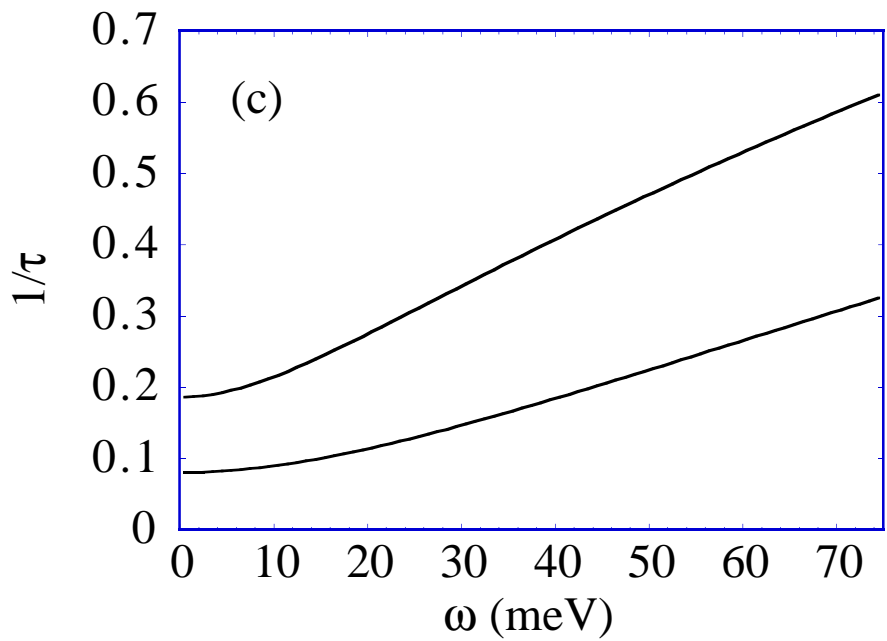
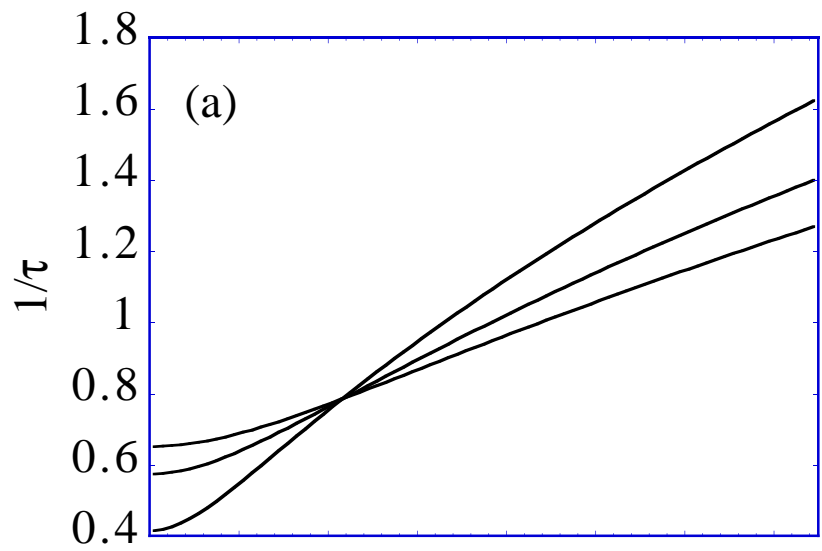
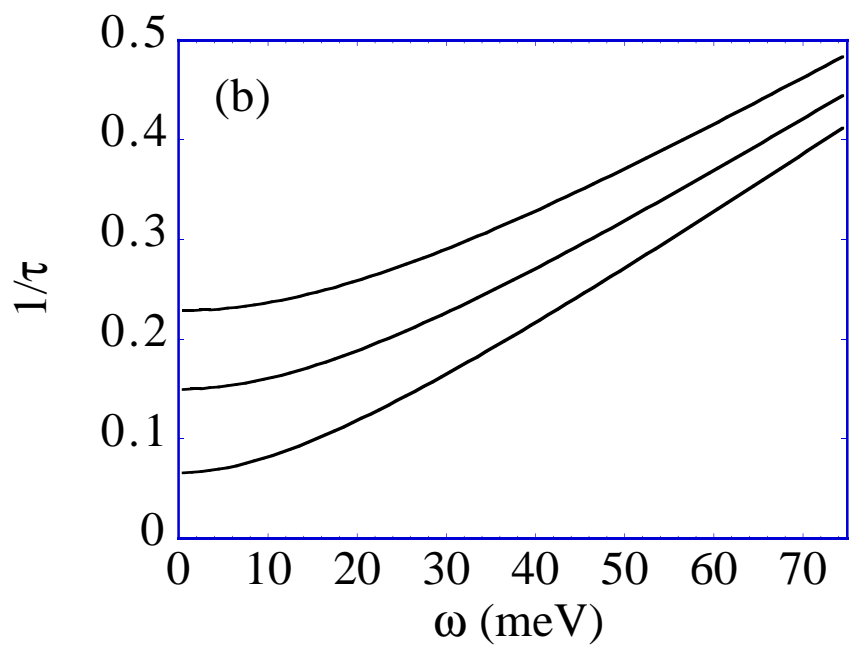
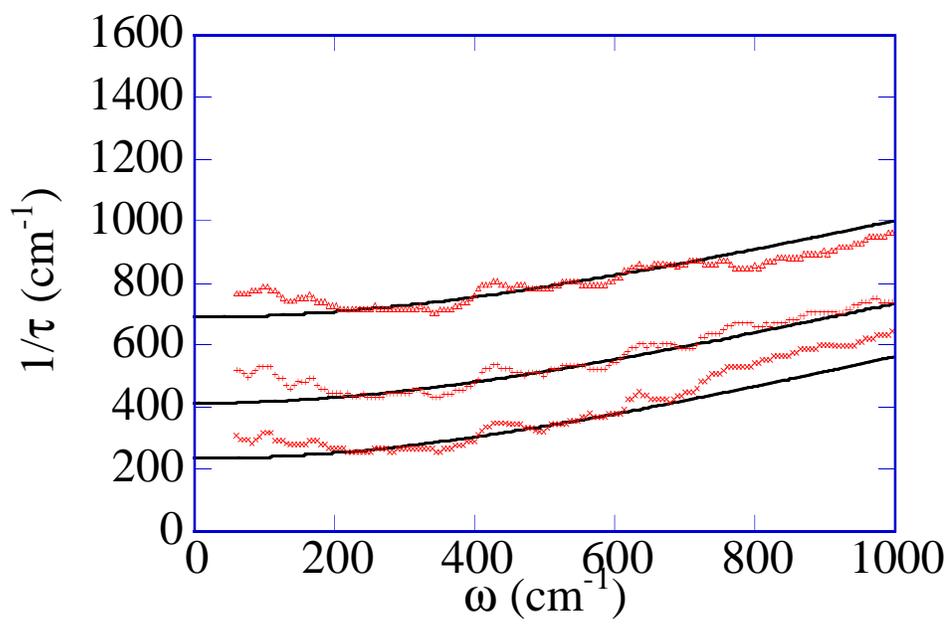
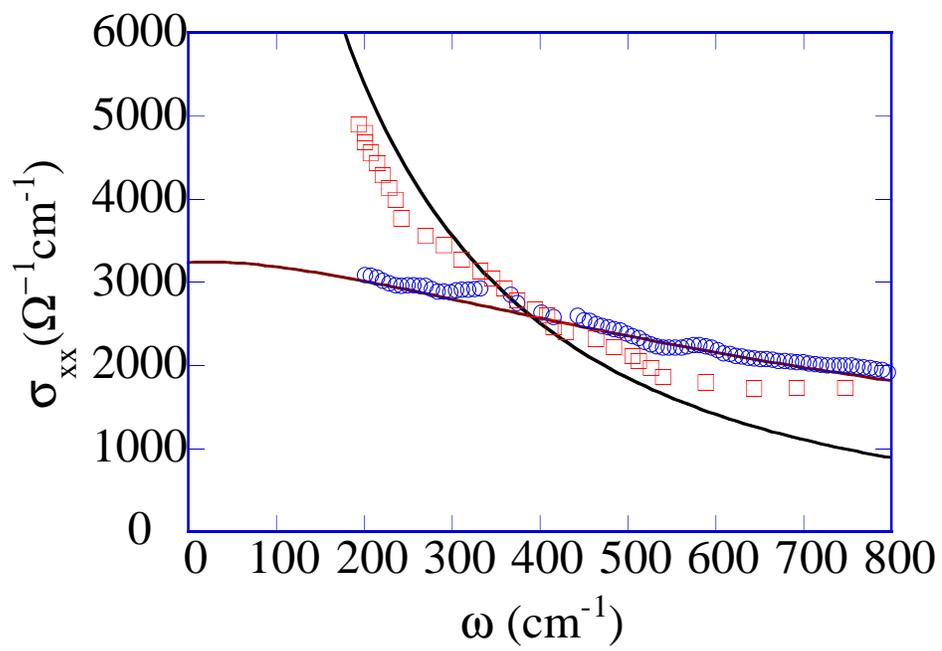
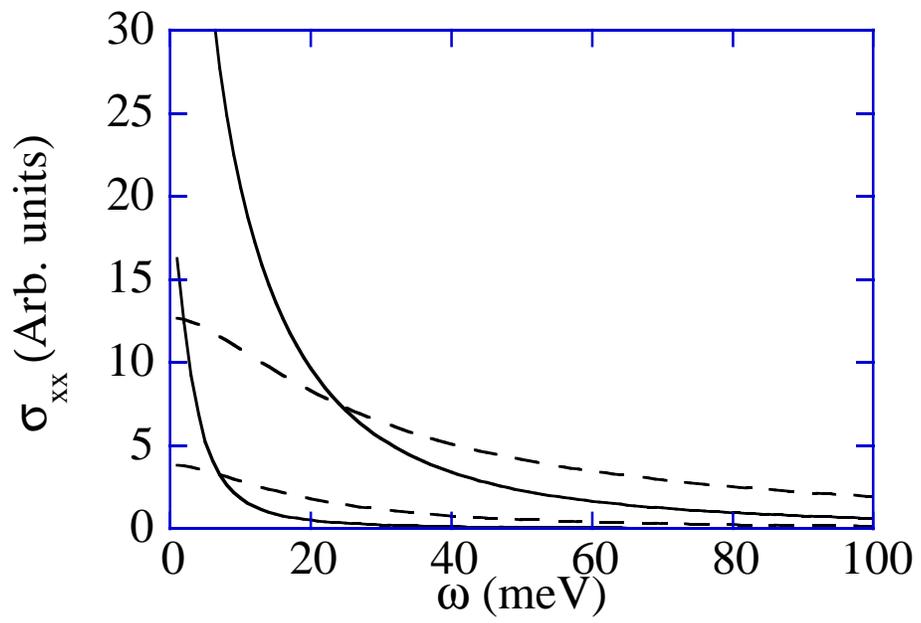
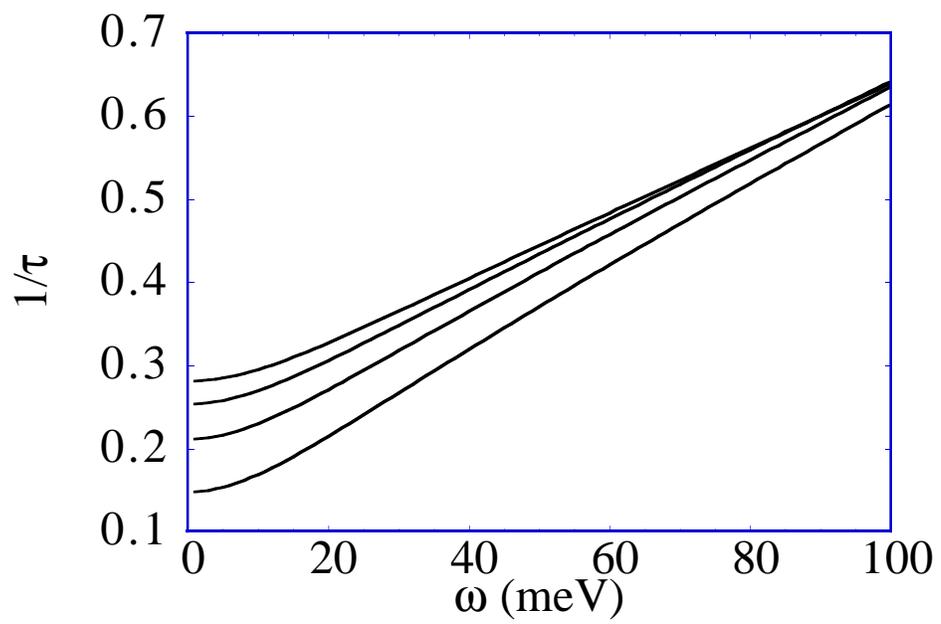
\end{document}